\documentclass{article}

\usepackage{PRIMEarxiv}

\usepackage[utf8]{inputenc} 
\usepackage[T1]{fontenc}    
\usepackage{hyperref}       
\usepackage{url}            
\usepackage{booktabs}       
\usepackage{amsfonts}       
\usepackage{nicefrac}       
\usepackage{microtype}      
\usepackage[numbers]{natbib}
\usepackage{fancyhdr}       
\usepackage{graphicx}       
\graphicspath{{media/}}     

\title{Who's actually being Studied? A Call for Population Analysis in Software Engineering Research
}

\author{
  Jefferson Seide Molléri \\
  Kristiania University College \\
  Kirkegata 24-26, Oslo - NO, 0153\\
  \texttt{jefferson.molleri@gmail.com} \\
}

\begin{document}
\maketitle

\begin{abstract}
  Population analysis is crucial for ensuring that empirical software engineering (ESE) research is representative and its findings are valid. Yet, there is a persistent gap between sampling processes and the holistic examination of populations, which this position paper addresses. We explore the challenges ranging from analysing populations of individual software engineers to organizations and projects. We discuss the interplay between generalizability and transferability and advocate for appropriate population frames. We also present a compelling case for improved population analysis aiming to enhance the empirical rigor and external validity of ESE research.
\end{abstract}


\section{Introduction}

In software engineering (SE) research, the practice of sampling is well-established, with many guidelines and experience reports for selecting representative subsets from diverse populations, e.g. \cite{baltes2022sampling, de2015investigating, kitchenham2002preliminary}. Researchers rely on sampling techniques, such as stratified and random sampling, to ensure their findings are representative of broader populations. However, without a proper characterization of the given population, the question highlighted in the title of this article - \textit{'Who's actually being studied?'} is left unaddressed, casting doubt on the representativeness of our findings. This position papers is driven by the need to bridge the current sampling practices with an comprehensive analysis of target populations in ESE research.

Population is defined as \textit{the complete set of entities that a researcher aims to study or understand} \cite{salkind2010encyclopedia}. Note that this is often misconceived as overly representative, and ideally it should be refined by the objective of the research. For instance, a research investigating the role of the Scrum Master is representative to software projects implementing the SCRUM methodology, not to all existing software projects. This concept is here referred as the 'target population,' which denotes a specific set of entities that the research is intended to generalize its conclusions to.

It is also vital to distinguish between generalizable and transferable results. Generalization concerns how findings apply to the target population, whereas transferability is about the relevance in comparable settings beyond that specific population. For instance, the findings of a study focused on Java projects could be generalized to all Java projects, and they might also be transferable to similar projects in other programming languages. Both generalizability and transferability are essential concepts for evolving SE research. 

Recently, empirical studies investigating issues related to practitioners' behavior and characteristics of software projects using GitHub became common. These studies (e.g. \cite{bissyande2013got, onoue2013study}) often state that is not possible to generalize the findings outside the context of the platform. However, they sometimes fail to generalize even within the GitHub environment due to insufficient population analysis. If the population is properly described, it is left to the reader to determine the applicability of these findings to their own practice.

Various methodologies incorporate sampling and population analysis. For instance, experiments require random assignment to ensure representative population characteristics among experimental groups. Surveys gather data from a broader population, depending on sampling frames to ensure representativeness. Case studies employ purposive sampling to select cases relevant to a targeted population. These methodologies are instructed by specific guidelines, such as \cite{wohlin2012experimentation, de2015investigating, runeson2012case}, emphasizing the importance of defining an suitable population frame and sampling strategy.

\section{Challenges and Considerations}

This section examines four distinct population types common in SE research: individual software engineers, organizations, software projects, and software artifacts. Each category presents unique challenges in terms of data availability, definition, and representativeness.

\subsection{Population Analysis for Individuals} 

Estimating the number of software engineers is challenging due to the lack of a comprehensive census, although estimates \cite{evans_worldwide_2023, statista_global_2023} are available. It's important to note that while these figures are crucial, the calculation of a minimum sample size for sufficient statistical power in research depends on more than just population size. Factors such as the effect size, which is the magnitude of the outcome being measured, and the significance level, which is the criterion for statistical significance are also vital \cite{dybaa2006systematic}. Additionally, information on average salaries and vacancies from work-related organizations (such as LinkedIn and Glassdoor) may help refining these estimations. They serve as indicators of demand and supply and help in establishing correlations with industry growth trends.


In addition to the population size, we lack understanding about the professionals' experiences and competencies. A notable study comparing novice software developers with students \cite{salman2015students} found just minor differences in their performance when applying a technique for the first time. By analysing student data in this context, we are able to project a future population frame of novice software developers, offering insights for understanding the evolving landscape of SE expertise. 

It is also important to reconsider how we distinguish between `professionals' and `students.' The traditional view of these as distinct groups is misleading, with many individuals embodying both categories, such as Master’s students working at software companies or open-source projects. This calls for a more nuanced analysis of population characteristics, moving beyond the simplistic classification to a more diverse landscape of SE expertise. This approach would help addressing the challenge of effectively measuring experiences and competencies specific to different roles in the field.

Another challenge arises when investigating the preferences in development practices, programming languages, and coding standards, as there are limited information available for accurately profiling the target population. Annual surveys from digital communities such as GitHub\footnote{\url{https://octoverse.github.com/}} and StackOverflow\footnote{\url{https://survey.stackoverflow.co/2023/}} provide demographic data. However, these surveys are constrained by community boundaries and self-selection bias, which may underrepresent the broader population of software developers.

The success of sampling and stratification techniques are severely limited by difficulties in accurately characterizing the population. While we assume diversity among software developers, we cannot substantiate this beyond specific contexts. The broader population, which is not easily identifiable, poses additional challenges. Nonetheless, it remains crucial to attempt defining the broader population, including its hidden segments, to ensure transferability, and potentially generalizability, of our findings.

\subsection{Population Analysis for Organizations} 

Analysing organizations, while similar to individuals, relies on estimating numbers of entities within the public and private sectors. National and regional surveys\footnote{See e.g. \url{https://economy-finance.ec.europa.eu/economic-forecast-and-surveys/business-and-consumer-surveys_en}} offer some data categorized by industrial sectors. We can use such reports to draw estimates for economically similar contexts. Note that global estimates are still challenging, as many less developed countries do not report such numbers.

In addition to this, the definition of an 'organization' in the context is often ambiguous. Many non-software companies have internal units or departments dedicated to develop and maintain software systems, like those in the automotive industry. These units are not counted within the software development sector in standard reports. Estimating the number of software development organizations within these companies is complex, still some data are available.

Another challenge arises in the study of software development teams. The composition and size of software teams could differ significantly, especially considering cross-functional teams and shared responsibilities. In such cases, characterizing the target population should focus on aspects such as culture, structure, and processes, to facilitate relevant comparisons.

Organizational population analysis is needed for industry segmentation and meaningful cross-company comparisons. Rather than focusing on numerical estimates, it is suggested to describe the contextual factors that helps us to understand the circumstances in which the phenomenon of investigation applies \cite{petersen2009context}.

\subsection{Population Analysis for Software Projects} 

Many software engineering studies focus on the characteristics of software projects. While these studies often using probabilistic sampling strategies aiming for representativeness and the ability to generalize conclusions, there is a critical aspect that needs attention: diversity. Accurate generalization depends not only on the sample size, but on a comprehensive understanding of the entire range of characteristics and variations present within the target population.

In reporting project characteristics, researchers must decide which aspects are key for drawing meaningful conclusions. This includes, but is not limited to, (1) project size, (2) project complexity, (3) development process, (4) project duration, (5) technologies adopted, (6) application area, (7) success metrics, (8) dependencies, and (9) versioning. It's vital that the chosen metrics are standardized and relevant to the target audience.

Data on open-source projects is readily accessible, and often these projects share many characteristics with proprietary ones. However, basing findings solely on open-source or proprietary projects severely restrict the reflection of diversity in software projects. It is challenging to discuss representative results without a comprehensive analysis of the entire project population and its characteristics.

\subsection{Population Analysis for Software Artifacts} 

Many software engineering studies focus on the technical aspects, such as architectural and code quality, compliance with standards, system performance, security, usability, and other software metrics. When investigating a specific software project or portfolio within a given organization as population, selecting a representative sample is both achievable and advisable. Yet, finding from such a narrowed study typically have limited transferability to external contexts. To ensure that the results have meaningful implications, a precise description of the studied population's characteristics is  required.

The study's goals or the phenomenon under investigation also guide the population analysis. Take the evaluation of development and operations (DevOps) practices, for example. A researcher would analyse the so-called DORA metrics: deployment frequency, lead time for changes, mean time to recovery, and change failure rate. To accurately characterize this population, the researcher must outline the overall distribution of such metrics, making sure that any sampling (or comparative case) aligns with the distribution, thereby avoiding any biases of misrepresentation.

\section{Case for Improved Population Analysis}


To effectively address the population analysis challenges outlined in this paper, we propose a set of practices. These recommendations are not sequential steps, but rather good practices to adapt to your research's specific needs. Resources like Baltes and Ralph \cite{baltes2022sampling}, Kitchenham et al. \cite{kitchenham2002preliminary}, and Nagappan et al. \cite{nagappan2013diversity} offer additional guidelines on designing empirical studies, including aspects of population analysis. 

\textbf{1. Establish population definitions and boundaries:} A first step to identifying a population frame is a clear description of who or what constitutes the population in the study. This involves determining whether the focus is on individuals, organizations, projects, or artifacts. It is crucial to establish a targeted population by setting boundaries for these definitions, e.g. software developers in mobile application projects. The boundaries ensure they are directly aligned with the specific objectives of the research. It's also important to acknowledge that the analysis represents a snapshot of the population at a specific moment in time. 


\textbf{2. Identify existing population datasets:} Based on the population description, we can identify existing databases, such as government statistics, industry analyses, and social media platforms. Utilize data mining techniques to leverage datasets from professional networks like GitHub, Stack Overflow, or LinkedIn, ensuring that the data aligns with the established boundaries of the target population. Note that social media may offer a skewed view, representing only the behaviors and preferences of self-selected individuals, thus not offering a fully representative sample. Collaborations with industry partners can grant access to more specific datasets, enriching the diversity of the data.

\textbf{3. Expanding and diversifying population datasets:} To create a comprehensive population frame, relying on a single data source is often insufficient. Researchers may combine data from various sources, which involves compiling and analyzing information from multiple relevant databases. Utilizing data mining and multivariate analysis helps in understanding the relationships among variables across databases.

\textbf{4. Cross-referencing and dataset validation:} Mind that the information provided by these sources might conflict and need cross-validation. For instance, two demographic studies \cite{statista_global_2023, evans_worldwide_2023} estimated the overall population of software developers in 2023 to be 27.7 million and 26.3 million, respectively. The discrepancies in the estimates are likely linked to how one defines 'software developer' and whether this data is collected on individual or organizational level. To address these inconsistencies, validation with experts is advised. This allows researchers to investigate factors causing variations in estimates.

\textbf{5. Advanced sampling techniques:} When a target population is hard to pinpoint, researchers may start with a small sampling and use those to refer others, i.e. \textit{snowballing sampling}. While this technique may not be ideal for capturing broad populations, it is highly effective for exploring specific niches or subsets, such as 'tech leaders in SCRUM teams.' Snowballing is often time-consuming and demands rigorous assessment of data quality and performance, e.g. by means of bootstrap methods assessing stability and variability of estimates from smaller samples. 

In the case of heterogeneous populations, \textit{stratification and weighting} ensure all subgroups are proportionately represented. By apply demographic and geographic filters, researchers can extract subgroups (strata) within our population that could be compared against broad population estimates. This approach helps evaluating the representativeness of subgroups and understanding how demographic and geographic factors influence the target population. 

\textbf{6. Reporting and documenting the population:} A final step in population analysis is thorough reporting the population frame. This includes, but is not limited to, the size of the population (and subgroups), their expertise, the complexity of the projects, the preferred methods and tools. Some quantitative characteristics may be expressed in absolute numbers such as the size or number of entities in the population. Others require descriptive statistics (mean, range, standard deviation, etc.) to convey demographic, geographic and temporal aspects. Qualitative characteristics, such as skills, competencies, culture, beliefs and attitudes require a narrative description.

\section{Conclusion}

Software engineering research strives for rigorous application of research methodologies to draw meaningful insights, yet the unique characteristics of the target population often remain unexplored. This paper has highlighted the challenges in identifying populations. A recurring theme is defining a representative sample that reflects the diversity of a target population. We have also highlighted the risk of non-generalizable results and the limited transferability of findings. The path forward requires that us, ESE researchers, employ robust statistics and methodological guidelines to ensure that the populations we study are accurately depicted in our studies.

\bibliographystyle{unsrt}  
\bibliography{main}

\end{document}